\pdfoutput=1

\documentclass[acmsmall]{acmart} 
\renewcommand\footnotetextcopyrightpermission[1]{} 
\AtBeginDocument{%
  }
    
\usepackage{algorithm}
\usepackage{algorithmic}

\usepackage{utfsym}

\usepackage{tabularx} 
\usepackage{array}

\usepackage{multirow} 
\usepackage{booktabs} 
\usepackage{makecell} 
\usepackage{array} 
\usepackage{multirow} 

\usepackage{ragged2e} 

\usepackage{caption}
\usepackage{array} 
\usepackage{cellspace} 
\usepackage{array} 
\usepackage{cellspace} 
\usepackage{colortbl} 
\definecolor{lightgray}{gray}{0.8} 

\usepackage[utf8]{inputenc}
\usepackage{listings}
\usepackage{xcolor}

\lstdefinelanguage{Solidity}{
    keywords={contract, function, public, payable, mapping, address, uint256, msg, sender}, 
    keywordstyle=\color{cyan!80!black}\bfseries, 
    morestring=[b]", 
    stringstyle=\color{orange!80!black}, 
    morecomment=[l]{//}, 
    morecomment=[s]{/*}{*/}, 
    commentstyle=\color{gray!60}\itshape, 
    sensitive=true 
}

\begin{document}
\title{Decentralized COVID-19 Health System Leveraging Blockchain}

\author{Lingsheng Chen}
\authornotemark[1] 
\affiliation{%
  \institution{Hainan University}
  \city{Haikou}
  \country{China}}

\author{Shipeng Ye}
\authornote{These authors contributed equally to this work.}
\affiliation{%
  \institution{Hainan University}
  \city{Haikou}
  \country{China}}
\email{yeshipeng35@gmail.com}

\author{Xiaoqi Li}
\affiliation{%
  \institution{Hainan University}
  \city{Haikou}
  \country{China}}
\email{csxqli@ieee.org}

\renewcommand{\shortauthors}{Ye and Li}

\begin{abstract}
With the development of the Internet, the amount of data generated by the medical industry each year has grown exponentially. The Electronic Health Record (EHR) manages the electronic data generated during the user's treatment process. Typically, an EHR data manager belongs to a medical institution. This traditional centralized data management model has many unreasonable or inconvenient aspects, such as difficulties in data sharing, and it is hard to verify the authenticity and integrity of the data. The decentralized, non-forgeable, data unalterable and traceable features of blockchain are in line with the application requirements of EHR. This paper takes the most common COVID-19 as the application scenario and designs a COVID-19 health system based on blockchain, which has extensive research and application value. Considering that the public and transparent nature of blockchain violates the privacy requirements of some health data, in the system design stage, from the perspective of practical application, the data is divided into public data and private data according to its characteristics. For private data, data encryption methods are adopted to ensure data privacy. The searchable encryption technology is combined with blockchain technology to achieve the retrieval function of encrypted data. Then, the proxy re-encryption technology is used to realize authorized access to data. In the system implementation part, based on the Hyperledger Fabric architecture, some functions of the system design are realized, including data upload, retrieval of the latest data and historical data. According to the environment provided by the development architecture, Go language chaincode (smart contract) is written to implement the relevant system functions.
\end{abstract}



\maketitle
\fancyfoot{}
\pagestyle{plain} 

\section{Introduction}
In 2008, Bencong developed the Bitcoin system\cite{nakamoto2008bitcoin}. With the rising price of Bitcoin, the underlying technology foundation behind the system, blockchain technology, has gradually entered everyone's vision\cite{panda2023bitcoin}, and has gradually become a research hotspot in the academic community. Today, blockchain layout frequently appears in government decision documents. The application of blockchain technology has become a hot research direction at present and in the future. More personnel have been devoted to research on blockchain technology \cite{dong2023blockchain}. The development of blockchain based application systems is related to the development of the Internet field and the future economic development of countries and regions.

With the vigorous development of Internet technology and applications, all walks of life have become inseparable from the Internet, including the medical industry \cite{li2023key,hong2022analyze,zhou2022use}. From the early queuing for registration and consultation, waiting for paper results, to the current simple online booking and inquiry, the Internet has greatly facilitated the medical industry. While enjoying convenience, one can also discover many problems, among which it is worth noting the storage and use of patient medical record information. According to rough statistics and calculations, the total storage capacity of medical data generated so far has reached an astonishing EB level \cite{mahajan2023smart}. Especially in the past decade, the popularization of digitization has led to exponential growth of medical data, and the rapid development of medical informatization has also posed new challenges to medical information sharing\cite{dennis2016temporal}.

Nowadays, more and more hospitals choose to store medical record information in the form of the Internet. Electronic medical records are gradually replacing traditional paper medical records, which greatly facilitates the development of the medical industry. When a patient needs to see a doctor, he does not need to provide his own paper medical records, but only need to query the personal medical record information in the electronic medical record management system. Through electronic medical records (EHR), doctors can easily understand the patient's past illnesses and medical information, and can make a more detailed and comprehensive judgment of the condition \cite{murala2023medmetaverse}. The treatment efficiency and treatment effect achieved by relying on electronic medical records are unmatched by traditional medical records. Meanwhile, when treating severe infectious diseases, EHR can share data with other excellent teams to help patients receive more comprehensive and rapid treatment \cite{zhu2024sybil}.

However, due to the fact that traditional database systems are centralized systems and system data is fully controlled by relevant institutions \cite{kassen2022blockchain,schlatt2022designing,tyagi2023decentralized}, it means that data between different hospitals is not shared or is difficult to share for the medical and health system. In the actual operation process, when patients come to different medical institutions for treatment, they usually need to handle the sharing of medical records themselves. This goes back to the traditional solution, which is not what a complete electronic medical record management system hopes for. In addition, since electronic medical records are ultimately stored in medical institutions, the management and use of medical records are opaque to patients, which is also something we do not want to see. EHR stores patients' personal information and medical records. Once attacked, it can lead to the leakage of sensitive information, such as patient privacy, causing security risks and doctor-patient conflicts\cite{zhong2023sybil}.

With the development of information technology, it is inevitable that EHR will gradually replace traditional case management. Past data \cite{henry2016adoption} also shows that various types of hospitals are increasing the proportion of electronic storage of cases. Therefore, solving the above problems has a very important practical application value. As the core technology of Bitcoin, blockchain has the properties of decentralization, data immutability, traceability, unforgeability, programmability, etc \cite{gharavi2024post}. The use of blockchain technology can establish a large-scale decentralized information management system for data sharing among multiple hospitals, which can facilitate hospitals to quickly query the medical records of various types of patients and solve the problem of medical data silos between hospitals. In addition, by designing a blockchain system, patients can have greater control over the use and information rights of their own medical records, providing better protection for their private data, such as medical records. And due to the tamper-proof and traceable nature of blockchain data, it can ensure that patient information will not be maliciously tampered with \cite{zhang2022blockchain}. Thus, build a decentralized, secure, and tamper proof medical information system.

The main contributions of this study are as follows:
\begin{itemize}
\item \textbf{Designed and implemented the blockchain-based COVID-19 health system architecture:} We Utilize the decentralized and tamper-proof features of blockchain to address issues such as difficulty in sharing medical data and opaque privacy.
\item \textbf{Introducing searchable encryption and proxy re-encryption technologies:} Implement retrieval and authorized access to encrypted data to enhance the security and sharing capabilities of system's data.
\item \textbf{Complete system prototype development based on Hyperledger Fabric:} We have implemented core functions such as data upload and query, and verified the feasibility and practicality of the design scheme.
\end{itemize}

\section{Background}

\subsection{Cryptography Technology}

\subsubsection*{\textbf{Hash Algorithm}}
As a representative of blockchain, the Bitcoin system has not experienced any effective attacks at the system level since its inception, and its system security is evident. Among them, the security of the blockchain data layer mainly relies on cryptographic-related technologies, including hash algorithms, digital signatures, and encryption algorithms.
Hash algorithm is a one-way function mapping that outputs a fixed-length hash value for an input string of any length. The formula is shown by eq.~\eqref{eq: Hash Algorithm}.

\begin{equation}
H(x) = y
\label{eq: Hash Algorithm}
\end{equation}

Where x represents input data of any length, and y is a hash value of a specified length.

A large number of hash algorithms have been proposed so far, among which MD5 \cite{rivest1992md5} and SHA \cite{eastlake2001us} are widely used. Among them, MD5 is widely used for data encryption and protection due to its universal, stable, and fast algorithm. However, with the continuous development of cryptanalysis, MD5 has been discovered to construct collision methods, which have been identified as insecure hash algorithms \cite{wang2005cryptanalysis}. In addition, the SHA algorithm (SHA1) with a shorter length y has also been cracked. The widely used SHA algorithm today is SHA2, including SHA256, SHA512, etc.

There are the following requirements for the design of cryptographic hash functions:

\begin{itemize}

\item \textbf{Anti Collision.}It can be divided into weak anti collision and strong anti collision. The former requires that the input x cannot be found for a given hash value y, so that the hash result of x is equal to it; The latter requires that no different inputs be found, so that their hash structures are equal.

\item \textbf{Irreversibility.}It refers to the one-way nature of hash functions, making it difficult to reverse calculate the corresponding input value based on the output hash value.

\item \textbf{Input Sensitivity.}Small changes in input values can lead to vastly different hash results.
    
\end{itemize}

The Bitcoin system uses SHA256 as the hash algorithm, and internal transactions in blocks adopt the Merkle Tree data structure, as shown in Fig 1. For each transaction data block, it serves as the input of the hash function, calculates its hash result, and uses it as input to the upper layer block. This process is repeated until a root hash is generated and stored in the block header. Therefore, based on the anti-collision characteristics of hash functions, it is not feasible to find a new transaction data block that keeps the hash result unchanged, thus ensuring that the data on the blockchain is tamper-proof\cite{kong2024characterizing}.

\begin{figure}[h]
    \centering
    \includegraphics[width=1.0\textwidth]{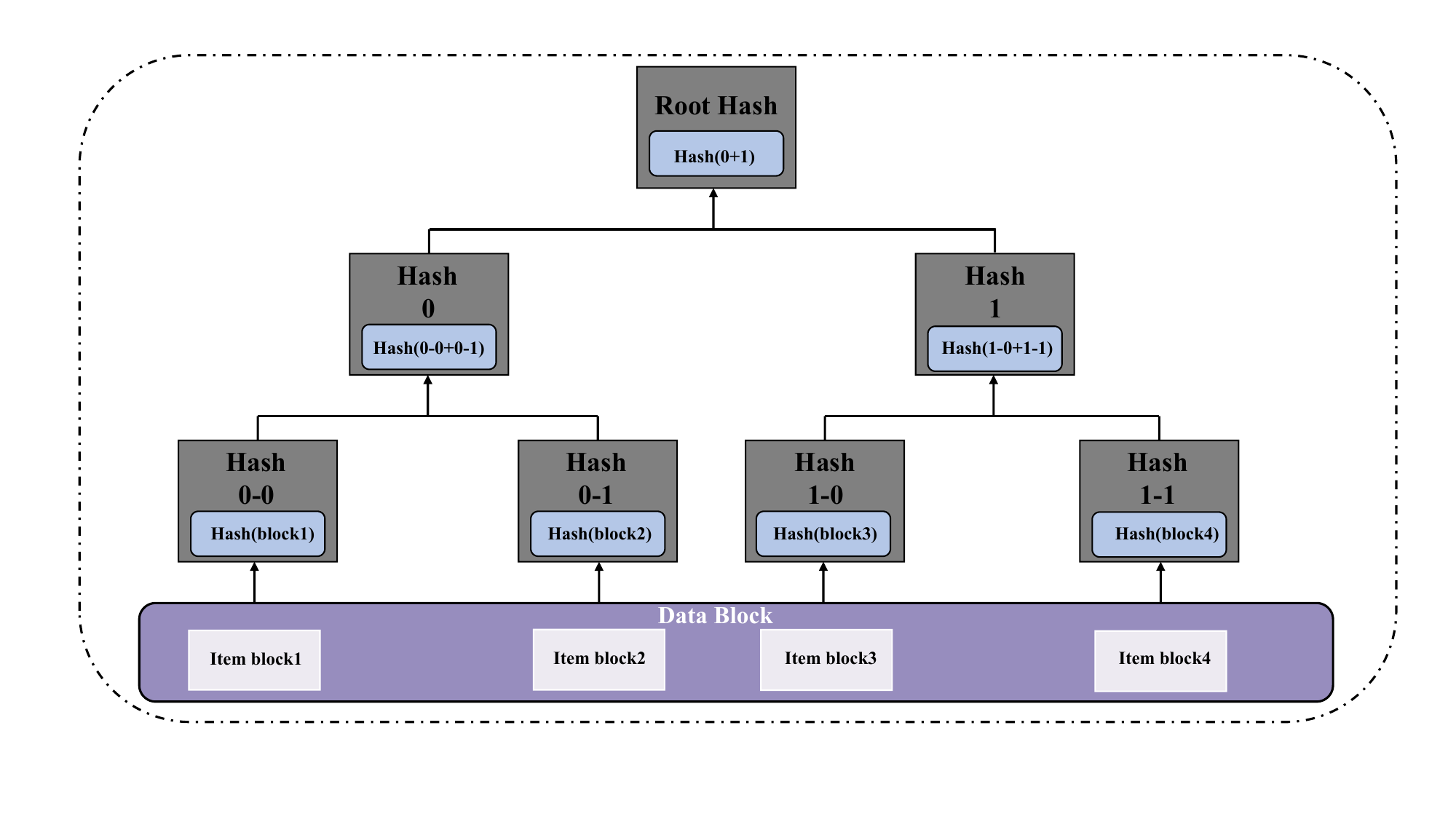}
    \caption{Merkle Tree Structure of Blockchain}
    \label{fig: Merkle tree structure of blockchain}
\end{figure}

In addition, research has shown that hash functions are puzzle friendly, meaning that the overall computation is memoryless. This characteristic is applied in the mining process of blockchain systems built on the POW mechanism, ensuring that miners' ability to obtain accounting rights is proportional to the computing power of nodes, and nodes with stronger computing power will not gain disproportionate advantages.

\subsubsection*{\textbf{Digital Signature Algorithm}}

A digital signature (also known as a public key digital signature) is a string of digits that cannot be forged by others and can only be generated by the sender of the information. This string is also an effective proof of the authenticity of the information sent by the sender. In 1977, the famous RSA algorithm was proposed, and modern cryptography developed the most important public key cryptography system \cite{imam2022effective}. It can be said that without the public key cryptography system represented by RSA, there would be no digital signature. To implement digital signatures, the following requirements must first be met :

\begin{itemize}

\item \textbf{Authentication.}Considering traditional symmetric encryption algorithms, encryption and decryption correspond to the signature and verification of digital signatures. However, since the private key owner is a single individual \cite{halak2022comparative}, it cannot meet the application scenarios of digital signatures. The emergence of public key cryptography allows signers to encrypt signatures with their own private keys, while validators can use publicly available signer public keys for decryption and verification, effectively verifying the signer's identity.

\item \textbf{Non-Repudiation.}After the signer signs the content, the characteristics of the signature algorithm and public key information cannot infer the private key data, which means that only the signer knows the private key \cite{liu2023idenmultisig}. For a signature that can be successfully verified by the public key, the signer cannot deny it.

\item \textbf{Data Integrity.}Due to the correlation between the input of the signature algorithm and the signed data file, this correlation can be achieved by directly using the file's data as input, or by using the file's summary information as input for operational efficiency \cite{hong2023graph}. This means that the information of the signed file cannot be modified.

\end{itemize}

Taking Bitcoin as an example, the Bitcoin system uses the elliptic curve-based digital signature algorithm ECDSA \cite{ulla2025research,huynh2023efficient,gani2024enhancing}. When a node publishes a transfer message, it first needs to sign the transaction information with a private key. After receiving the transaction information, the miner node verifies the transaction data signature with the signer's public key. The characteristics of digital signature algorithms ensure the verifiability, integrity, and non-repudiation of transaction data.

\subsection{Consensus Mechanism}

As blockchain is a decentralized system, for public chain systems, each system node can independently choose to join this distributed system. Nodes communicate with each other through a P2P network, and each node is jointly responsible for the normal operation of the blockchain system. For such a complex distributed network system, in order to enable nodes to reach a consensus and jointly build the blockchain system they belong to, the design of the blockchain consensus algorithm is extremely crucial \cite{bu2025enhancing}.

Since the concept of blockchain was proposed, consensus algorithms have always been a key research direction in blockchain technology. Among them, typical consensus algorithms include the Proof of Work (POW) mechanism, the Proof of Stake (POS) mechanism, the Delegated Proof of Stake (DPOS) mechanism \cite{larimer2014delegated}, and the Practical Byzantine Fault Tolerance (PBFT) algorithm\cite{castro1999practical}.

The PoW algorithm originates from Bitcoin. The core idea of this algorithm is to allow nodes to compete for the right to record transactions by competing in computing power. To make nodes willing to serve the blockchain system, the system sets up corresponding incentive mechanisms, rewarding nodes that obtain the right to record transactions with a certain amount of tokens. In the Bitcoin system, it is stipulated that the hash value of the block header must be less than a set value. Due to the characteristics of the hash algorithm, nodes can only continuously modify the random number and keep trying to find a block whose hash value is less than the difficulty value \cite{ali2023proposal}, and then broadcast the corresponding block for verification by other nodes.

The PoS algorithm was proposed based on the PoW algorithm. In response to the huge energy consumption problem caused by the computing power competition in the PoW algorithm's mining process, PoS adopts the approach of granting the right to record transactions to the node with the highest stake in the system \cite{rahman2024proof,fior2024innbc,feng2022regulatable}. This stake can be reflected in the blockchain system of cryptocurrencies as the quantity and holding time of tokens held by the node. This means that nodes with a larger number of tokens and longer holding periods have a higher possibility of obtaining the right to record transactions \cite{mivsic2023toward}. The DPoS algorithm is a further development of the PoS algorithm \cite{selvam2024development}. Its core idea is to use a voting election method to select nodes with the right to record transactions. First, N representative nodes need to be elected, and these representative nodes take turns to obtain the right to record transactions. According to the actual operation, representative nodes can be re-elected.

The BFT algorithm\cite{lamport2019byzantine} takes into account the possibility of malicious nodes in the system. When the number of Byzantine nodes is less than one-third of the total number of nodes, nodes can reach a consensus normally. Representative algorithms of the BFT algorithm include the PBFT algorithm, which mainly optimizes the efficiency of the BFT. The PBFT algorithm includes three stages: pre-preparation, preparation, and commit. Since nodes need to communicate with each other, the communication complexity of the system increases with an increasing number of nodes, and the communication cost is relatively high. Therefore, this algorithm is not suitable for public chain scenarios with a large number of nodes, but is suitable for consortium chains and private chains.

\subsection{Smart Contract}

In 1997, cryptographer Nick Szabo first proposed the concept of smart contracts\cite{zou2025malicious}, where Szabo hoped to eliminate intermediaries and make contracts automatically effective as long as certain conditions were met.

The emergence of blockchain technology has driven the development of smart contracts. No need for a third party, the contract is a computer protocol that exists in code form \cite{sharma2023review,afraz2023blockchain,ahmed2024smart}. Its essence is a computer program written by programmers and stored in blocks, using deterministic algorithms and data sources, and satisfying terminability. When the network state changes and meets the initial conditions set, the 'miner' will automatically execute the smart contract content. In Ethereum, smart contracts are considered programs that can directly control digital assets. Ethereum and Hyperledger Fabric are two representative technology platforms for applying smart contracts.

\subsection{Searchable Encryption Technology}

When storing information on an untrusted server, such as email servers and file servers, we hope to protect the stored data in an encrypted form to reduce security and privacy risks \cite{gupta2022secure}. But such measures taken for safety will cause us to lose some of our original functions. For example, if a client wants to retrieve its stored files through keywords, ordinary encrypted data algorithms cannot match the ciphertext with the given search term, making it impossible to achieve the corresponding search function. Therefore, in the past, it was not known how to make the data storage server perform searches and answer queries without losing data confidentiality \cite{wu2022ensure}.

When users need to store tamper-proof data, they can use blockchain systems to achieve this. The data on the blockchain is open and transparent, and any node has the right to view the content on the chain. Therefore, before users store data, they need to encrypt it. And when data needs to be used later and it is desired to retrieve data through keywords, applying searchable encryption technology is the only option. Combining blockchain with searchable encryption technology can provide data confidentiality without changing the practicality of the system, and has good application prospects.

\subsection{Access Control Technology for Permission}

When users need to store private data in the blockchain system and retain retrieval functionality, searchable encryption technology can be used. For encrypted data, there may be application scenarios that need to be shared. For example, the COVID-19 Health System studied in this paper needs to provide information sharing to doctors when considering medical treatment due to infection. There are currently two feasible sharing strategies: one is for users to decrypt the retrieved data and share the plaintext data; The second is to enable data sharers to independently access data information through certain permission access control technologies, including proxy re-encryption technology and attribute encryption technology.

The concept of proxy re-encryption was first proposed by BLAZE et al.\cite{blaze1998divertible}, and its core idea is to use a proxy to perform secondary encryption on stored encrypted data. The result of the secondary encryption can be decrypted by the private key of the data's sharer, and the proxy cannot obtain any plaintext information during the process \cite{lang2014cryptographic}.

The process of proxy re-encryption is as follows:

\begin{enumerate}

\item Each node user has generated their own public and private key data. Assuming that the data sharing parties and the public and private key data are Alice: Sk1, Pk1; Bob: Sk2, Pk2.The proxy node is a server.

\item Alice first encrypts the ciphertext data with her own private key Sk1. Bob sends his public key to Alice to request data sharing rights.

\item Alice generates a data delkey based on Bob's public key Pk2 and her own private key Sk1, and sends it to the server. The server re-encrypts the ciphertext using the delkey and sends the result to Bob. At this point, Bob can decrypt the data using her own private key.

\end{enumerate}

\section{System Requirements and Analysis}

\subsection{System Architecture Design}

\subsubsection{\textbf{Overall Architecture}}

As shown in Fig 2, the overall architecture of the COVID-19 Health System in this paper, in which the system composition can be divided into six parts, and the specific introduction of each part is as follows:

\begin{itemize}

\item \textbf{Certificate Authority (CA)}: Responsible for issuing certificates to various system components except BC, generating and distributing all public and private key data required for DO and DU, managing the use of key and certificate data for the entire system, including functions such as generation, distribution, update, audit, revocation, and destruction.

\item \textbf{Data Owner (DO)}: The data owner is an ordinary user who is mainly responsible for uploading their own data, retrieving their own data when using it, encrypting private data specifically, and setting access policies for data that needs to be shared.

\item \textbf{Data User (DU)}: It is an object that DO needs to share data with, and requests access to data from DO. After completion, it can query the shared data files through MIFS.

\item \textbf{Medical Institution Alliance Server Cluster (MIFS)}: The entire system is responsible for responding to various requests from DO and DU, and completes queries and flexible storage of user data by accessing DDBS and BC, and completes data sharing according to access policies.

\item \textbf{Distributed Database System (DDBS)}: Used to coordinate on-chain and off-chain storage to share the storage pressure on the blockchain.

\item \textbf{Alliance Blockchain (BC)}: Store the data or data summaries submitted by DO on chain, and implement system functions based on the smart contracts deployed on the blockchain system.
    
\end{itemize}

\begin{figure}[h]
    \centering
    \includegraphics[width=1.0\textwidth]{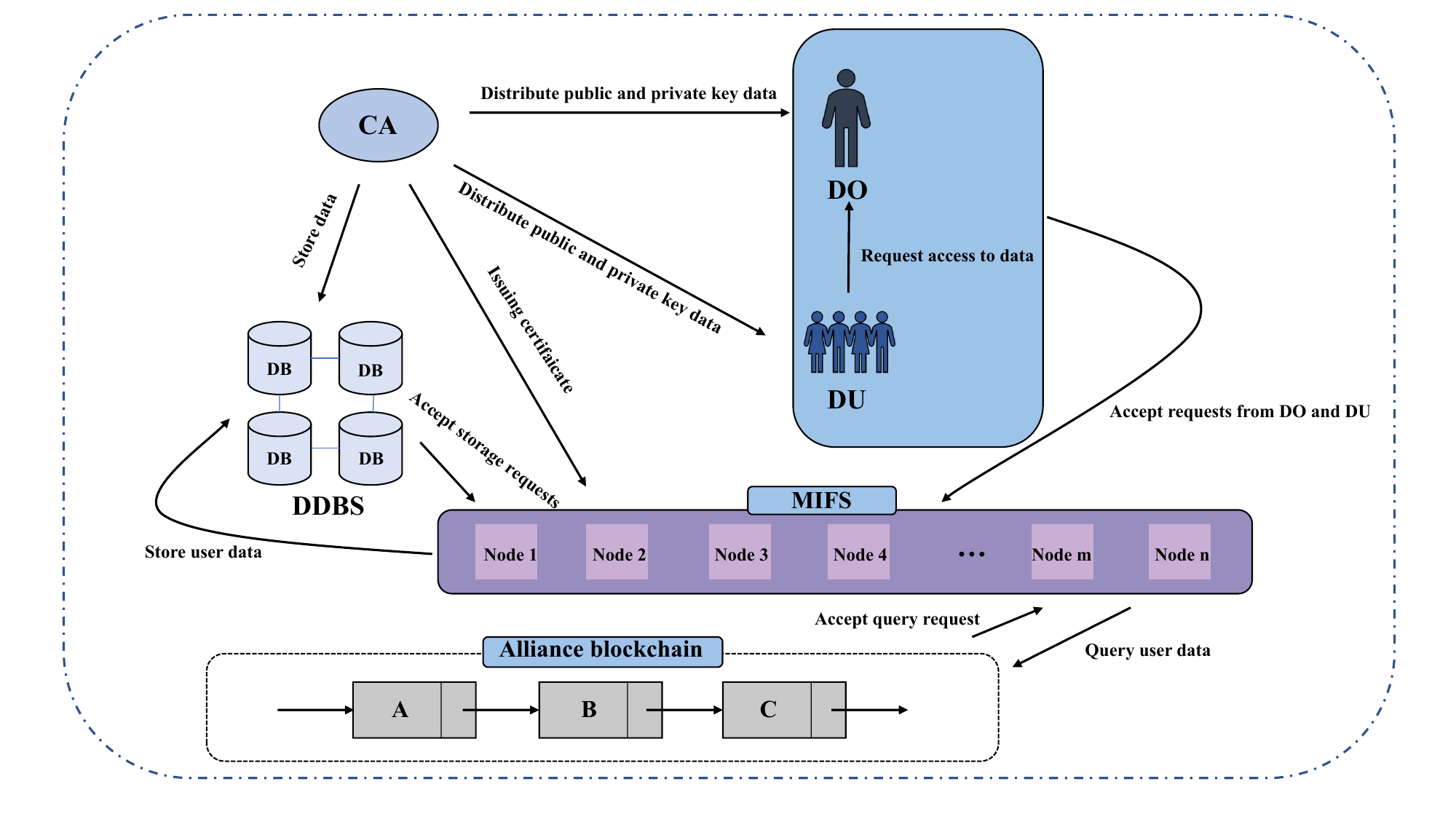}
    \caption{Logical Architecture of COVID-19 Health System}
    \label{fig: Logical architecture of COVID-19 Health System}
\end{figure}

\subsubsection{\textbf{Running Process}}

The specific implementation process of the blockchain system is as follows:

\begin{enumerate}

\item CA Performs Global Parameter Setting

\begin{itemize}

\item \textbf{Issue Certificates}: At the beginning of the system, the CA is MIFS, DDBS issues certificates, and DO and DU are two types of users in the system. They need to submit a request to join the system to the CA, which reviews the application information. If they meet the requirements, the CA issues certificates to the users and adds them to the blockchain system.

\item \textbf{Key Generation}: Based on the requirements of DO and DU, CA can generate the keys needed for DO and DU, and manage the entire key usage process through the organization's principles of master key, secondary key, and tertiary key. In addition, considering the possible issue of users' lack of trust in the security of CA management keys, support for user clients to generate keys locally, provide public key information registration, and subsequently identify identity through certificates \cite{li2024detecting}.

\end{itemize}

\item DO Submission and Access Data

\begin{itemize}

\item \textbf{DO Upload Data}: Based on the amount of data submitted, different considerations are made. For small data, MIFS chooses to store it directly in BC, while for large file data, MIFS stores it in DDBS. The data content is organized through the Merkle tree storage structure, and the root hash value is finally used as the data digest information for on-chain storage.

\item \textbf{DO Accesses Data}: Related to storage, there are also two scenarios for DO to access data. For small data access, MIFS completes data queries by calling smart contracts deployed on BC based on specific access requests and returns them to DO. For large confidential data, MIFS retrieves DDBS through searchable encryption methods and returns data results. In order to facilitate subsequent file sharing, a proxy re-encrypted copy can be stored separately. In addition, an additional function can be added for data queries, where DO and DU can request the Merkle path of the data from MIFS to verify the authenticity of the returned data.

\end{itemize}

\item DU Accesses Data

As the data sharer, DU needs to first request DO to access the specified data. When DO agrees to DU's data access request, subsequent operations are carried out.

\begin{itemize}

\item \textbf{Unencrypted Data Access}: Unencrypted data is marked with the public key information of the data owner to indicate the source of the data. When DU cannot access it normally, DO returns its signature file after agreeing to the access request. DU submits the corresponding data access request to MIFS with the DO's permission. Furthermore, set the start and end times of sharing in the signed license file to achieve temporary permission access.

\item \textbf{Encrypt Data Access}: For shared access to encrypted data, the DO can first save a ciphertext file encrypted by proxy re-encryption technology for the encrypted file that needs to be shared when uploading the file. After receiving the access request from DU, DO generates a re-encryption key based on DU's public key and sends the permission access signature file and proxy re-encryption key to DU. When DU accesses data, it sends a signed file and a re-encryption key to MIFS. After successful verification, MIFS returns the re-encrypted file content.

\end{itemize}

\item MIFS Data Management

MIFS is the key to processing information in this system architecture, responsible for multiple request operations, including DO and DU, and is also the only system architecture that can directly access DDBS and BC.

\end{enumerate}

\subsection{Functional Requirement}

\subsubsection{\textbf{Universal Module}}

The general function module includes various general functions of system users, which have similar implementation processes. Considering the convenience of later maintenance and program scalability, functions with similar functions or implementation processes can be grouped into one module\cite{ting2017system}.

The general functions mainly include key management, new user registration, user login, and other aspects. Considering the high degree of overlap between the functions of general modules and the CA institutions in the system architecture, modules will be classified into CA institutions, and the introduction is equivalent to the introduction of CA's functions in this blockchain system.

User General Function Analysis:

\begin{enumerate}

\item User Registration

New user registration is divided into DO registration, DU registration, MIFS registration, and DDBS registration. The registration of MIFS and DDBS requires administrator review, so a different interface is needed specifically for medical institution registration and review. After successful registration, it needs to work continuously in the system to support the composition and operation of the system. This part of the registration generally needs to be set in the network environment initialization during the initial stage of system establishment.

When registering, users need to choose their identity, whether they are a patient or a doctor, and enter their real name, phone number, and password. The doctor also needs to fill in the code of the medical institution they belong to, indicating their hospital of employment. The phone number is the unique identifier for users in the system and cannot be duplicated. It is used for users to log in after registration.

\item User Login

Users need to log in to the system to see the specific operation interface, which is divided into DO login and DU login. The user needs to enter the basic information required for login and select the identity type for login. The system will automatically determine whether the user has entered the information and registered. After successful login, users will be redirected to the corresponding operation interface for different identity types. If unsuccessful, an error message will be returned.

\item Key Generation

In the process of user registration and key update, a key needs to be generated. The key generation module uses the internal random number generator of the system to generate a key bit string with good randomness, and calculates additional system parameters through the RSA algorithm and selected system parameters. Key generation, as a part of system calls, is also the most important part of key management to support system operation. The key generation function is described in Table 1.

\begin{table}[h]
\centering
\caption{Key Generation}
\label{tab:key_generation}
\begin{tabular}{@{}ll@{}}
\toprule
\textbf{Function Name}   & Key Generation \\
\textbf{Function Role}   & DO, DU \\
\textbf{Function Description} & Invoked during User Registration and Key Update \\
\textbf{Input}           & Invocation Information \\
\textbf{Output}          & Randomly Generated Secure Key \\
\bottomrule
\end{tabular}
\end{table}

\item Key Update

Users may lose their private keys due to negligence or other unexpected circumstances while using the system; Or due to the long usage time of the private key, there may be serious security risks; Or the user wishes to modify the key. It is necessary to abandon the use of the original key and update the key accordingly. The user key update function is shown in Table 2.

\begin{table}[h]
\centering
\caption{Key Update}
\label{tab:key_update}
\begin{tabular}{@{}ll@{}}
\toprule
\textbf{Function Name}   & Key Update \\
\textbf{Function Role}   & DO, DU \\
\textbf{Function Description} & Generate New Key Pairs for User Use \\
\textbf{Input}           & User ID \\
\textbf{Output}          & New Key Pair \\
\bottomrule
\end{tabular}
\end{table}

\item Key Audit

The entire usage cycle of the key needs to be audited to understand the usage of the key, effectively evaluate and analyze the security of the key, and prevent possible malicious use after the key is stolen. The key audit function is shown in Table 3.

\begin{table}[h]
\centering
\caption{Key Audit}
\label{tab:key_update}
\begin{tabular}{@{}ll@{}}
\toprule
\textbf{Function Name}   & Key Update \\
\textbf{Function Role}   & CA \\
\textbf{Function Description} & Real-time Audit of Managed Key \\
\textbf{Input}           & Secret Key \\
\textbf{Output}          & Key That May Pose Security Risk \\
\bottomrule
\end{tabular}
\end{table}

\end{enumerate}

\subsubsection{\textbf{Data Owner Module}}

COVID-19 health data belongs to the privacy of DO. The design of this system needs to ensure that DO has absolute control over the data. As stipulated at the beginning of this chapter, all data is divided into public data and private data according to whether the information is encrypted or not. For public data, the access control mechanism of the system is set up, which is not public to DU. When DU wants to access public data, it also needs to apply for access permission from DO.

This section introduces the functional requirements of the DO module, which mainly includes login and registration, data upload and query, authorization sharing, and key management. The uploading, querying, and sharing of data are basic functional requirements for data management. Depending on the requirements for data confidentiality, there are significant differences in the implementation of these functions, which need to be considered separately. The design of the key management module is aimed at the decentralization of key management, supporting users to independently manage their keys in the local key management module.

\begin{enumerate}

\item Key Management

For the keys used by users, two management strategies are provided. In addition to being selected by the user during system registration, subsequent change operations are also provided. For the two management strategies, one is to entrust the CA organization to manage the keys, and achieve key management by calling the key management function of the module; The second is to manage and access keys offline through a locally set key management module.

\item Data Upload

According to the type of uploaded data, it can be divided into public data upload and private data upload. For public data, encryption is not required and will be processed on the blockchain after being handed over to MIFS. Before uploading private data, DO needs to perform pre-encryption processing, which can be further divided into searchable encrypted data and re-encrypted data according to the requirements of their usage scenarios. In fact, the corresponding original data is the same, only encrypted using different encryption methods and stored to support subsequent data query functions.

\item Data Query

DO queries can be divided into public data queries and private data queries based on the differences in the types of data being queried. For public data, which is on-chain data, it is implemented through smart contracts deployed on the blockchain. In this design, two query functions are considered: one is to query the latest data, which can obtain the latest health code, nucleic acid test results, and other information; the other is to query historical data, and the query results return all on-chain data information of DO in chronological order. For privacy data stored in a distributed database, it is stored in a format that requires a searchable encrypted ciphertext.

\item Authorized Sharing

When authorizing shared data, DO divides it into public data authorization and private data authorization based on the different types of data. After verifying the identity of the applicant, DO chooses whether to proceed with shared authorization. For the authorization of public data, DO needs to return a signature file that specifies key information such as shared data content, authorization start time, and authorized party's public key data; For the authorization of private data, DO also needs to return a file signed by its own private key, but the difference is that a record about the re encryption key needs to be attached to the file for DU to query.

\end{enumerate}

\subsubsection{\textbf{Data User Module}}

The data user module is a functional module related to DU. Compared to DO, DU cannot upload data files. This module consists of four main parts: registration and login, key management, request sharing, and data query \cite{morris2025trusted}. The function of the key management submodule is equivalent to the key management function introduced in DO. Due to DU's inability to upload data, the data source for DU's data queries comes entirely from request sharing, and the results that can be queried exist in the dataset that has received the sharing request.

\begin{enumerate}

\item Request Sharing

Requesting sharing is the only way for DU to obtain data sources, and after a successful request, it will receive a DO-signed permission to access the file. Before handing over the request data to DO, DU needs to initiate request sharing, call the request sharing module, standardize the file information requested for sharing and its own identity proof, and generate and send them.

\item Data Query

The data query process of DU is similar to that of DO, with two differences. Firstly, when DU initiates a query, it must submit a file that has been authorized for the query. During the query process, the authorized file will be compared, and the query content must not exceed the range set specified in the authorized file. The second is to search for keywords based on searchable encrypted data when querying private data, and the returned results of the query conform to the ciphertext data structure of proxy re-encryption.

\end{enumerate}

\subsubsection{\textbf{Alliance Server Group Module}}

The MIFS module is the only system component in the system that can directly communicate with BC and DDBS. All upload and query requests related to data from system users need to be processed by MIFS, so MIFS is an indispensable component of the system. The main functions of MIFS include data storage and querying.

\begin{enumerate}

\item Data Storage

MIFS can be divided into public data storage and private data storage when storing data. For the storage of public data, it is the same as data storage on ordinary blockchain, directly stored on the chain in the structure of Merkle Tree; For the storage of privacy data, due to its large capacity, a combination of on chain and off chain storage is used, with DDBS storing the main data and on chain storage of summary information (tree roots).

\item Data Query

The introduction of the data query function refers to the introduction of the data query in DO and DU. MIFS only serves as a relay for querying and completes the real query operation according to the user's request.

\end{enumerate}

\section{System Design and Implementation Details}

The system implementation part of this article is based on the Hyperledger Fabric architecture, as shown in Fig 3. The detailed system implementation can be roughly divided into three parts, namely the deployment of the Fabric network, SDK instantiation, and front-end design\cite{yu2024toward}. The Fabric network mainly includes node definition and chain code installation. As the middle layer structure of the system, the SDK is responsible for deploying the system network downwards, conveying upper-layer operation instructions, and returning instruction execution results upwards. Front-end programs provide services through a web format.

\begin{figure}[h]
    \centering
    \includegraphics[width=1.0\textwidth]{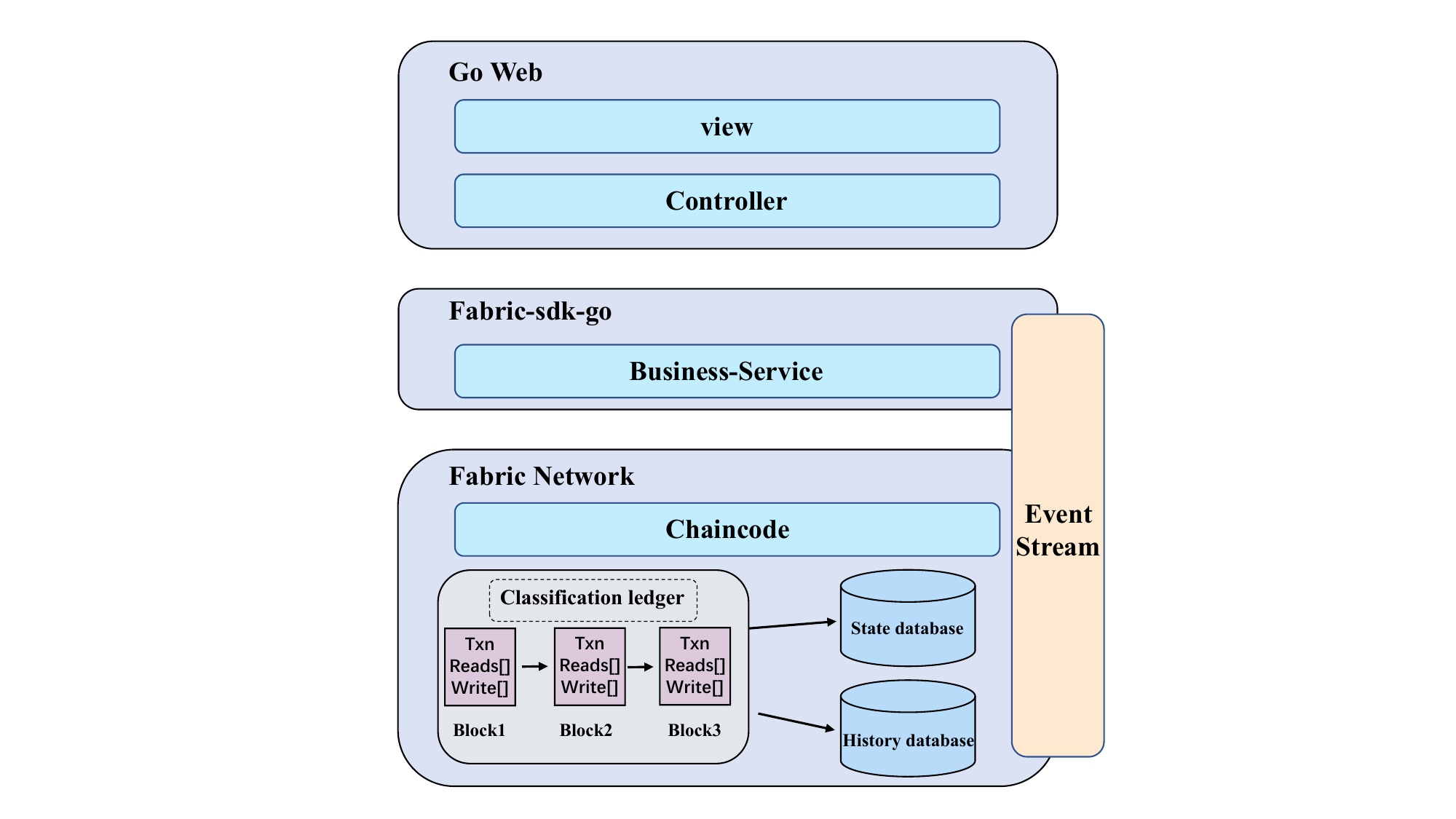}
    \caption{System Architecture Diagram}
    \label{fig: System Architecture Diagram}
\end{figure}

\subsection{Hyperlegder Fabric}

\subsubsection{\textbf{Project Overview}}

Hyperledger (or Hyperledger Project) is an open-source project aimed at promoting the cross-industry application of blockchain. It was initiated by the Linux Foundation in December 2015, and its members include leaders in the finance, banking, IoT, supply chain, manufacturing, and technology industries. This project was created to promote further development of the blockchain industry and facilitate cooperation between various blockchain projects. This project will inherit independent open protocols and standards, using framework methods and dedicated modules, including consensus mechanisms and storage methods for each blockchain, as well as identity services, access control, and smart contracts \cite{mao2024scla}.

Hyperledger Fabric is an excellent representative of consortium chains\cite{androulaki2018hyperledger} and a platform that provides distributed ledger solutions. Each architecture in Fabric is modular and supports the combination of different components, allowing for the centralized processing of independent modules and the construction of an actual executable system.

According to the design concept of the system architecture, the Fabric architecture has proposed some new concepts, such as the concept of channel, which is an independent hyperledger fabric instance. This is the first concept related to blockchain systems proposed in Fabric. Channel is similar to the concept of a subnet in a network, where all channels are independent and do not depend on each other. Different channels have different rules and cannot share data with each other. The chain code mentioned earlier is closely connected to the channel during deployment, meaning that different channels need to deploy different chain codes \cite{zhang2009channel}.

\subsubsection{\textbf{Project Workflow}}

The Fabric blockchain defines three types of nodes, namely: Client (SDK), Peers, and Orderers. The client is similar to a client node and can make various requests to the system. The peers node records the ledger information in the blockchain system, while the orderers are similar to miners in the public chain, responsible for sorting and managing the update requests of the ledger and notifying the peers node to update the ledger. The three of them work together to form the entire blockchain network of Hyperledger Fabric\cite{foschini2020hyperledger}. The workflow of Fabric is shown in Fig. 4.

\begin{figure}[h]
    \centering
    \includegraphics[width=1.0\textwidth]{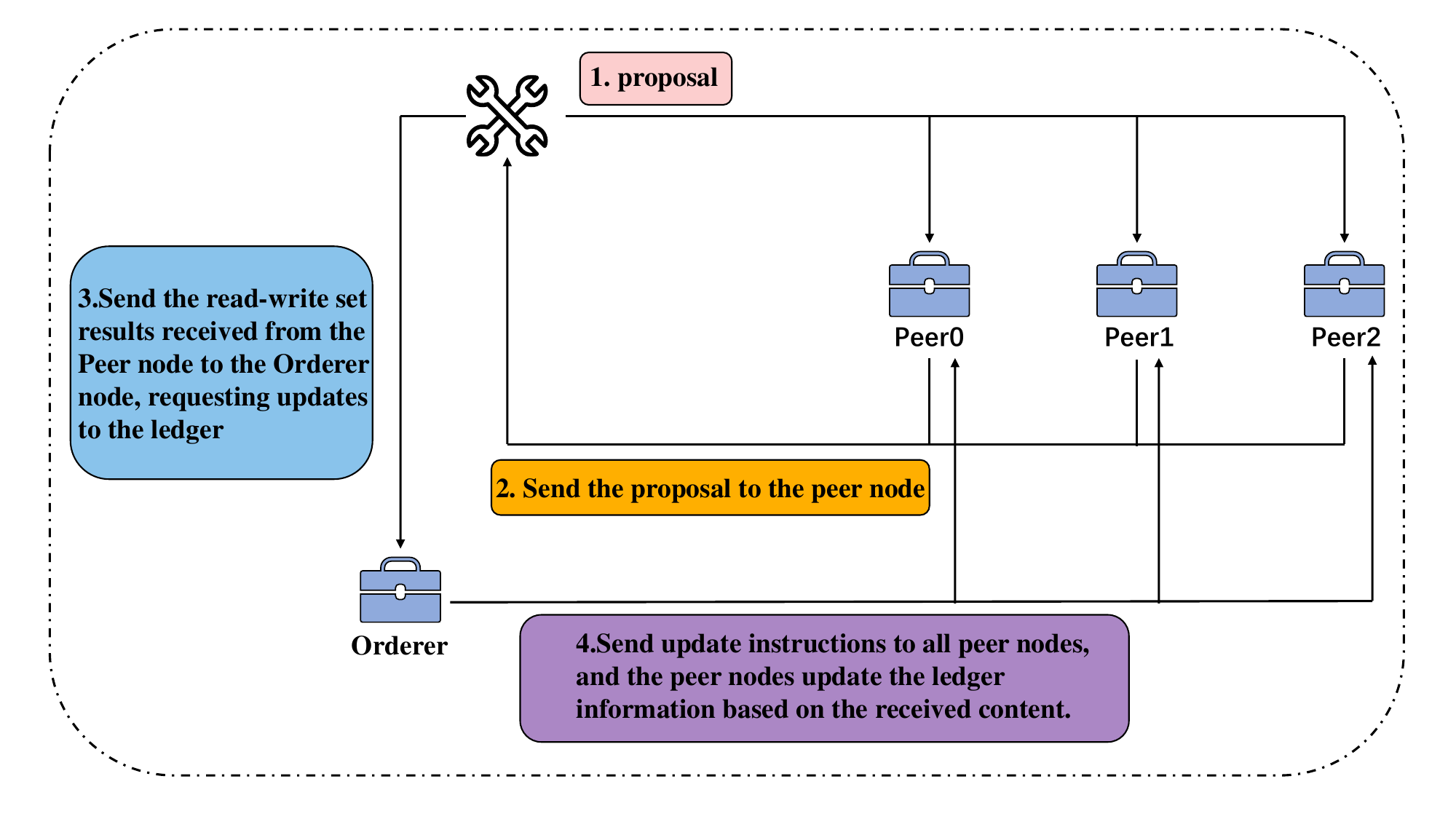}
    \caption{Fabric workflow}
    \label{fig: Fabric workflow}
\end{figure}

Proposal operation is when the SDK wants to update the ledger data; it first needs to initiate a proposal that specifies the data that needs to be updated, and the proposal needs to be sent to the Peer nodes that exist in the system. Then the Peer node will simulate the execution of the received proposal content based on the current ledger version. The execution result is a generated read-write set, which records the specific update operations and the updated ledger version. After the simulation execution is completed, the Peer node will sign the running result and release it to the SDK to complete the endorsement operation.

In the update request operation, the SDK sends the read-write set results received from the Peer node to the Orderer node, requesting updates to the ledger. After receiving a certain amount of data, the Orderer node will process the data and first consider sorting the information. Its main purpose is to prevent the double-spending problem in the blockchain system. Then, it is necessary to check whether the digital signatures of the Peer nodes are correct and whether the simulated execution results of each Peer, that is, whether the read-write set is consistent, are checked. After all detections are passed, the Sorting node will send update instructions to all Peer nodes, and Peer nodes will update the ledger information based on the received content\cite{li2024defitail}.

The sorting operation of the Orderer node is performed after receiving a certain number of data requests or after a certain period of time. The Orderer node sorts the received requests and packages them into data blocks of a certain format, then broadcasts the new blocks to the Peer node. Sorting services require consensus to ensure that all members receive a set of transactions in the same order, maintaining data consistency.

\subsection{Configuration of Blockchain Network Environment}

\subsubsection{\textbf{Channel Division}}

This is an advanced Fabric concept, as Fabric is a distributed system that typically involves multiple organizations, and different versions of Fabric code may (and typically) exist on different nodes within the network and channels within the network. Fabric allows this without requiring every peer node and sorting node to be at the same version level. In fact, supporting different version levels is the reason for supporting structural node rolling upgrades\cite{zhang2017android}.

Channel, as a new concept in Fabric, is necessary for network deployment. In the implementation process, this article simplifies the division of channels, as shown in Fig.5, by considering adding all Peer nodes to the same channel. In fact, in practical implementation, peer nodes can be assigned to different channels to identify the node's affiliation due to the different needs of the actual organization. Nodes on different channels can communicate by setting anchor nodes. A more specific consideration is that when a medical institution does not want to share some user data, or when users do not want to have on chain public data, a separate channel can be set up to complete small-scale internal communication.

\begin{figure}[h]
    \centering
    \includegraphics[width=1.0\textwidth]{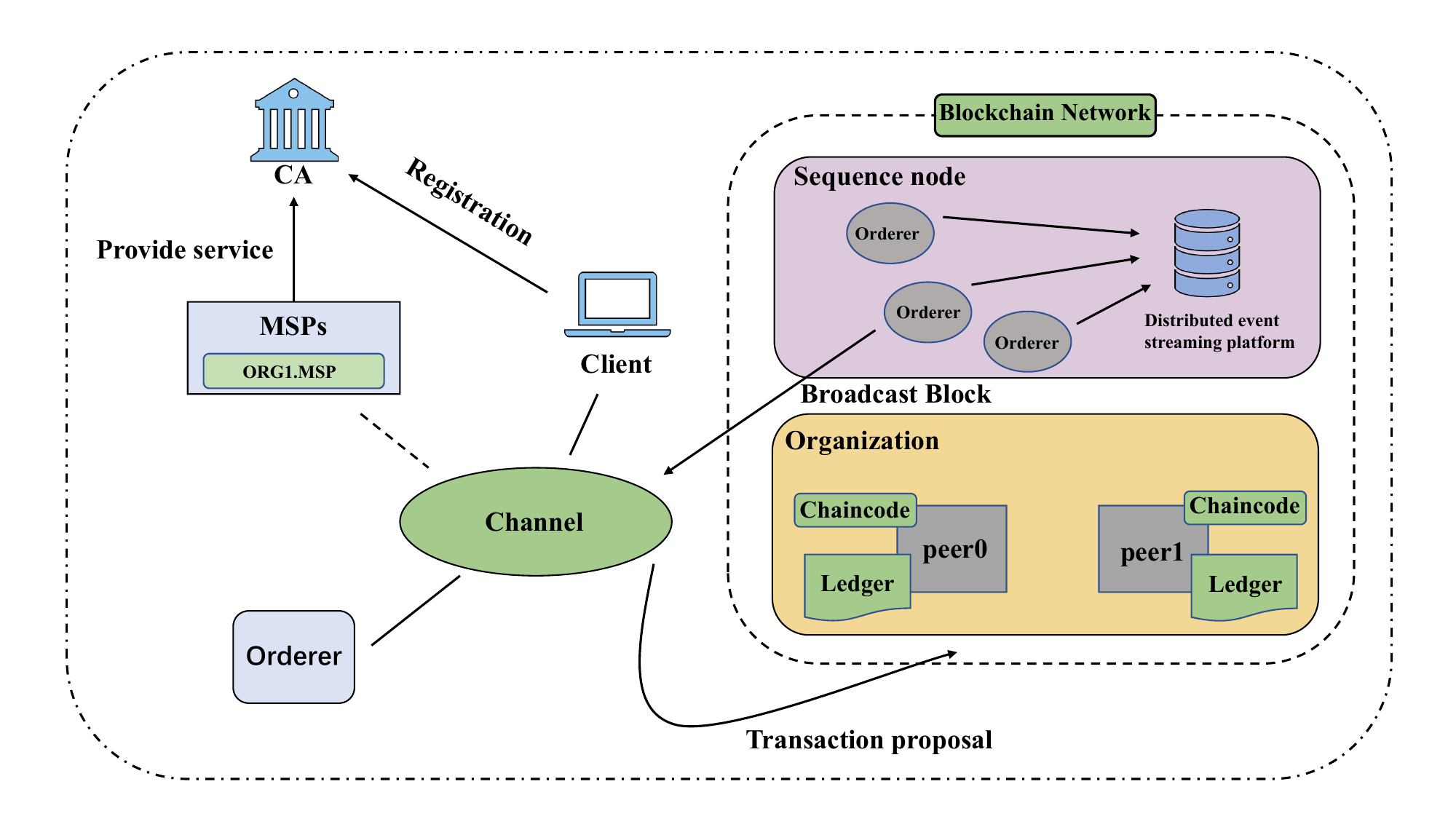}
    \caption{Fabric Network Topology Diagram}
    \label{fig: Fabric network topology diagram}
\end{figure}

\subsubsection{\textbf{Peer and Orderer Node Deployment}}

When building a blockchain environment for system operation, we can selectively create information about peer and orderer nodes. According to Listing 1, the detailed configuration code of the relevant nodes is as follows:

\begin{lstlisting}[caption={The Detailed Configuration Code of The Relevant Node}]
OrdererOrgs:
  - Name: Orderer
    Domain: example.com
    EnableNodeOUs: true
    Specs:
      - Hostname: orderer
PeerOrgs:
  - Name: Org1
    Domain: org1.example.com
    EnableNodeOUs: true
    Template:
      Count: 2
    Users:
      Count: 1
\end{lstlisting}

An organization can contain multiple nodes and users, and a user can have multiple nodes. The above code defines an orderer node and an org organization containing two peers, with Domain identifying the respective domain names. Two Peer nodes belong to one user, and it is evident that one user does not possess decentralized characteristics. Therefore, this is only a rough definition for experimental testing.

After defining the information of the relevant nodes, the system functions in the Fabric architecture can be called to generate, including: MSPs generating certificate proofs for all nodes through CA, generating public and private key information for nodes, etc.

After the deployment of the relevant nodes is completed, a genesis block needs to be created, which is stored in the orderer node \cite{antal2021distributed}. Various information about the nodes needs to be written into the block to ensure the integrity of the data. All defined nodes are irreplaceable and tamper-proof. Once the data is uploaded, it is fixed, and others cannot impersonate a valid node or modify the content of the node, thus ensuring the security of the node data information and reducing the impact on the entire deployment. The creation of the genesis block is contained in the configtx.yaml folder. In addition to the information mentioned above being written into the block, it also defines some important parameter information about the blockchain system, including the sorting mode of sorting nodes, block generation time and capacity restrictions, specific code. The example is as follows:

\begin{lstlisting} [caption={The Example of Specific Code}]
OrdererType: solo
BatchTimeout: 2s
BatchSize:
    MaxMessageCount: 10
    AbsoluteMaxBytes: 99 MB
PreferredMaxBytes: 512 KB
\end{lstlisting}

The sorting mode is defined as solo here. Similarly, due to the limited number of nodes in the experiment and the simple network topology, solo mode has been used. For complex network deployments, Kafka mode can be used instead.

The BatchTimeout parameter defines the block generation time of the blockchain system, where blocks are generated every two seconds. The MaxMessage Count parameter is the maximum number of requests that the orderer node is likely to receive while waiting. 10 means that when 10 pending requests are received, they need to be processed immediately, even if the waiting time has not ended. Absolute Maxbytes and Preferred Maxbytes, respectively, specify the maximum and minimum capacity of a single block, with a maximum capacity of 99MB and a minimum capacity of 512 KB.

In addition, the file also contains information for creating channels. After running, the channel will be created, and nodes will be added to it.

\subsubsection{\textbf{Chain Code Design and Deployment}}

For the design process of a system based on the Hyperledger Fabric architecture, because the chain code is the only means of modifying the blockchain ledger, the design of the chain code is also the most important part. This is related to all the business functions that the designed system needs to complete, and it is also the key to distinguishing batches of different blockchain systems.

When formally writing the chain code operation, additional consideration needs to be given to the ID card information, including when storing data. Due to the need to query historical data based on the ID card information in the future, the ID card needs to be used as a key value, and the information table needs to be stored as a value. Based on this, write the go function for data upload as shown in Listing 3.

\begin{lstlisting}[caption={Core Code for Data Upload}]
func (t *EducationChaincode) addEdu(stub shim.ChaincodeStubInterface, args []string) peer.Response {
    if len(args) != 2 {
        return shim.Error("The number of given parameters does not meet the requirements")
    }
    var edu Education
    err := json.Unmarshal([]byte(args[0]), &edu)
    if err != nil {
        return shim.Error(" An error occurred during information deserialization") 
    }
    // Duplicate check: The ID number must be unique
    _, exist := GetEduInfo(stub, edu.EntityID)
    if exist {
        return shim.Error("The ID number to be added already exists")  
    }
    _, bl := PutEdu(stub, edu)
    if!bl {
        return shim.Error("An error occurred while saving the information")  
    }
    err = stub.SetEvent(args[1], []byte{})
    if err != nil {
        return shim.Error(err.Error())
    }
    return shim.Success([]byte("Information added successfully")) 
}

\end{lstlisting}

The function first performs compliance checks on the basic parameters of the input and then performs deserialization operations on the input. Before depositing, it is necessary to verify the compliance of the ID card information, that is, to determine whether it duplicates the existing ID card. After all is correct, the storage will be carried out with the ID card as the key value.

Listing 4 shows the core code for updating data. Similarly, the number of input parameters for the function is checked first, and then the input data is deserialized \cite{liu2024serdesniffer}. Then check if the changed information is compliant, that is, the ID card information cannot be changed. After all checks are completed, assign all updated data values to complete the update operation.

Listing 5 shows the query operation for the latest data. The query is completed based on the name and certificate number, and the number of input parameters is checked to see if it is two. Before querying, the input is concatenated into a query string called queryString. When querying the input, the function getEduByqueryString written to query the data table based on the string, is called to complete the query. After the query is correct, the query result is returned.

\begin{lstlisting}[caption={Core Code for Data Update}]
func (t *EducationChaincode) updateEdu(stub shim.ChaincodeStubInterface, args []string) peer.Response {
    if len(args) != 2 {
        return shim.Error("The number of given parameters does not meet the requirements") 
    }
    var info Education
    err := json.Unmarshal([]byte(args[0]), &info)
    if err != nil {
        return shim.Error("Failed to deserialize edu information")  
    }
    // Query information according to the ID number
    result, bl := GetEduInfo(stub, info.EntityID)
    if!bl {
        return shim.Error("An error occurred while querying information according to the ID number")
    }
    result.Name = info.Name
    result.BirthDay = info.BirthDay
   ...
    _, bl = PutEdu(stub, result)
    if!bl {
        return shim.Error("An error occurred while saving the information")
    }
    err = stub.SetEvent(args[1], []byte{})
    if err != nil {
        return shim.Error(err.Error())
    }
    return shim.Success([]byte("Information updated successfully"))
}

\end{lstlisting}

The general process of data query based on ID card number is also similar. The difference is that the function called by the query is more ID card-based and needs to query historical data. It calls the historical data query function stub. GetHistoryForKey under the development architecture \cite{li2024cobra}.

For the designed chain code, it needs to be deployed to the peer nodes of the system. Otherwise, even nodes in the same channel and organization cannot synchronize data with nodes in the system due to the inability to update ledger information through the chain code. Essentially, they do not belong to nodes in the blockchain. Therefore, for both peer nodes in this system, the chain code needs to be deployed. The deployment process includes packaging and installation of the chain code, organization recognition of the chain code, checking whether the chain code is ready and defining and initializing operations before officially running the system. The specific deployment refers to the CreateCCLifecycle function in sdkInit/sdkSetting.

\begin{lstlisting}[caption={Core code for Latest Data Query}]
func (t *EducationChaincode) queryEduByCertNoAndName(stub shim.ChaincodeStubInterface, args []string) peer.Response {
    if len(args) != 2 {
        return shim.Error("The number of given parameters does not meet the requirements")
    }
    CertNo := args[0]
    name := args[1]
    // Assemble the query string required by CouchDB (a standard JSON string)
    // queryString
    queryString := fmt.Sprintf("{\"selector\": {\"docType\": \"eduObj\", \"CertNo\": \"%s\"}}", CertNo)
    queryString = fmt.Sprintf("{\"selector\": {\"docType\": \"%s\", \"CertNo\": \"%s\", \"Name\": \"%s\"}}", DOC_TYPE, CertNo, name)
    // Query data
    result, err := getEduByQueryString(stub, queryString)
    if err != nil {
        return shim.Error("An error occurred while querying information by certificate number and name")
    }
    if result == nil {
        return shim.Error("No relevant information was found according to the specified certificate number and name")
    }
    return shim.Success(result)
}

\end{lstlisting}

\subsection{System Network Service Startup}

The system testing in this paper was conducted in a virtual machine and ran smoothly in the Ubuntu 20.04.3 environment. As it is based on the Hyperledger Fabric architecture, it is necessary to prepare the environment according to the official documentation before testing, which includes the installation of Docker, Docker Comspose, and Golang. Additionally, since the system runs locally, Docker Accelerator and Go Agent need to be configured. The shell code for configuring the Go Agent is as follows:

\begin{lstlisting}[caption={The Shell Code for Configuring The Go Agent}]
go env -w GO111MODULE=on
go env -w GOPROXY=https://goproxy.cn,direct
\end{lstlisting}

After the above environment configuration is completed, the following configurations need to be made before running:

\begin{enumerate}

\item Set the GOPATH path. Set the GOPATH environment variable of the GO language to the src folder path where the program is located. For example, if a file is pulled to the/root/go/src path, the configuration statement for the environment variable is export GOPATH=/root/go.

\item Add system dependencies. The dependency information of the Go language is saved in the go.mod file. Enter the program folder and enter the go mod tidy command to add dependencies to the system \cite{wang2024smart}.

\item Execute the running script. Executing the ./ clean-dacker.sh file to complete system startup.

The above clean-dacker.sh file contains the following content:

\begin{lstlisting}[caption={The Content Contained in The clean-dacker.sh file}]
sudo docker rm -f $(sudo docker ps -aq)
sudo docker network prune
sudo docker volume prune
cd fixtures && docker-compose up -d
cd ..
rm education
go build
./education
\end{lstlisting}

The first three instructions are the cleaning work performed before running. As the Faric system module runs in different Docker containers, it is necessary to clean up the Docker container-related content before running, including deleting Docker container content, deleting unused networks, and deleting volumes that are not used by any container. The fourth step involves setting up the Faric network environment before system startup, which includes creating all nodes, creating channels, and packaging and deploying chain codes. The last three instructions perform the final execution work, and the executable file 'education' will be generated by running 'go bulk'. After running, the entire system is started and can be accessed through a browser. The defined access interface address here is 127.0.0.19000.

\end{enumerate}

\section{Conclusion}

This paper selects the current hot spot of COVID-19 as the application scenario, and designs the blockchain-based COVID-19 Health System. Its essence is to study the application of blockchain technology in the electronic medical system, in order to solve the problems of data isolation, easy to tamper with, and opaque to users of traditional electronic medical systems by applying the characteristics of blockchain technology, such as decentralization, openness, transparency, and non-tampering. The design phase of this article combines technologies such as blockchain, searchable encryption, and proxy re-encryption to create a more secure and powerful blockchain system. And based on the differences in privacy requirements of users' different data, the data is divided into public data and private data. Different data are designed with different data organization structures and storage methods. On the basis of ensuring that data advocates have absolute control over the data, users can manage their own data more flexibly and diversely. In terms of system implementation, satisfactory engineering implementation has been achieved for the basic management functions of blockchain-based data. Based on the development environment of the Hyperledger Fabric architecture, the blockchain network of the system is deployed. By writing and installing smart contracts for the system, the implemented functions include user login, data upload, latest data query, and historical data query. Finally, by writing front-end programs and SDK code, the system's functions were received and implemented through front-end calls and presented externally through web services.

\bibliography{main} %
\bibliographystyle{apalike}
\end{document}